\documentclass[useAMS,usenatbib,usegraphicx]{mn2e}

\usepackage[T1]{fontenc}
\usepackage{aecompl}

\title[Motion of BLR clouds]{On the orbital motion of cold clouds in BLRs}
\author[M. Shadmehri]{Mohsen Shadmehri$^{1,2}$\thanks{E-mail:
m.shadmehri@gu.ac.ir; mmshadmehri@gmail.com} \\
$^{1}$Department of Physics, Faculty of Sciences, Golestan University, Gorgan 49138-15739, Iran\\
$^{2}$ Research Institute for Astronomy and Astrophysics of Maragha (RIAAM), Maragha, Iran, P. O. Box: 55134-441}
\begin{document}

%\date{Accepted 1988 December 15. Received 1988 December 14; in original form 1988 October 11}

\pagerange{\pageref{firstpage}--\pageref{lastpage}} \pubyear{2015}

\maketitle

\label{firstpage}

\begin{abstract}
We study orbit of a pressure-confined cloud in the broad-line region (BLR) of active galactic nuclei (AGNs) when   the combined effects of the central gravity and anisotropic radiation pressure and the drag force are considered. Physical properties of the intercloud gas such as its pressure and dynamic viscosity  are defined as power-law functions of the radial distance. For a drag force proportional  to the relative velocity of a cloud and the background gas, a detailed analysis of the orbits is performed  for different values of the input parameters.  We also present analytical solutions for a situation where the intercloud pressure is uniform and the  viscosity  is proportional to the inverse square of the radial distance. Our analytical and numerical  solutions demonstrate decay of the orbits because of considering the drag force so that a cloud will eventually fall onto the central region after so called time-of-flight. We found that time-of-flight of a BLR cloud is proportional to the inverse of the dimensionless drag coefficient.  We discuss  if time-of-flight becomes shorter than the life time of the whole system, then  existence of mechanisms for continually forming BLR clouds is needed.
\end{abstract}
\begin{keywords}
galaxies: active - galaxies: nuclei
\end{keywords}

\section{Introduction}%\label{s:?}
In a unified theory for the structure of active galactic nuclei (AGNs), various parts have been proposed to explain the observational features of these interesting astrophysical objects \citep[e.g.,][]{netzerbook}. Existence of cold clouds embedded in an intercloud medium  known as broad-line region (BLR) may explain some of the observational evidences. Even near to the Galactic center, there are clouds in their orbits around the central supermassive  black hole \citep[e.g.,][]{Gill,burkert12}. Apparently coexistence of clouds with a hot gaseous intercloud medium is common in some of the accreting systems.
Current attempts to study these clumpy systems are concentrated on three main aspects. Understanding processes which may lead to the formation of BLR clouds is an active research field \citep[e.g.,][]{frome,pitt}.  Moreover, there are noticeable uncertainties about stability of these clouds and the  confinement mechanisms in the light of theoretical arguments and recent numerical simulations \citep{Rees,krause12,namekata}. Regardless of current uncertainty over the nature of the confinement mechanisms, orbital  motion  of the BLR clouds and their radiated emission allow us to estimate mass of the central black hole \citep[e.g.,][]{marconi,netzer2010}. Considering different forces that may affect orbit of a BLR cloud and the importance of the orbit's shape, some authors addressed the orbital analysis of the BLR clouds during recent years \citep[e.g.,][]{krause11,plewa,khajenabi15}.

To make progress in this field, orbital motion of a cloud is treated like a classical two-body problem where gravitational force of the central black hole and a force due to the radiation of an accretion disc, as the dominant forces,  are controlling   the net force on each cloud and the resulting orbits \citep[e.g.,][]{Liu11,krause11,krause12}. It is generally assumed that the clouds are optically thick. In another effort, just recently, a few authors studied dynamical motion of clouds through a gaseous medium semi-analytically by assuming that the ensemble of clouds is collisionless and obtaining solution of the corresponding Boltzmann equation under simplified conditions \citep{wang,mina}. Simplicity of treating the system as the two-body central force problem and its implications on calculating the broad-line emission, however, deserve further studies in analyzing orbital motion of BLR clouds in particular regarding to the unexplored physical ingredients. 

BLR clouds are subject to the anisotropic radiation pressure of a central source as has been pointed out by \cite{Liu11}.  Recently, a few authors studied orbit of BLR clouds under combined effects of the central gravity and anisotropic radiation pressure \citep{krause11,plewa,khajenabi15}. For pressure-confined BLR clouds with constant column density, \cite{plewa} presented analytical solutions for the orbit of the clouds. None of the previous studies of BLR cloud's orbit considered interaction of a cloud with the ambient medium in the form of a resistive force. Although effect of the drag force on the orbital motion of the BLR clouds has not been studied, in the context of planetary motion and orbital analysis of the satellites in a central gravitational force, a number of authors treated the classical two-body problem with a resistive force and presented numerical or even analytical solutions \citep[e.g.,][]{humi,mitt,mavr}. 

\cite{humi} studied properties of the orbital motion of an object in a central force with a drag force opposite to the velocity vector and proportional to a power-law function of the magnitude of the velocity and the distance to the center of attraction. Based on their  solutions, they demonstrated decay of the orbits when the drag force is considered.  When the drag force is proportional to the velocity and the inverse square of the radial distance, \cite{mitt} obtained an analytical solution. Then, \cite{mavr} generalized the analysis to a case with an isotropic radiation force and found three integrals of motion. 

In this paper, we extend previous studies of BLR cloud's orbital motion by considering  a drag force which is opposite  to the velocity of a cloud. We prescribe physical properties of the intercloud gas such as its pressure and dynamic viscosity  as power-law functions of the radial distance. In the next section, general formulation of the problem is presented. In section 3,  analytical solutions are presented for a given set of input parameters. Then, a detailed analysis of shape of orbits is performed in section 4. We conclude with possible astrophysical implications of our results in section 5.

\begin{figure}%[tb]
\includegraphics[scale=0.35]{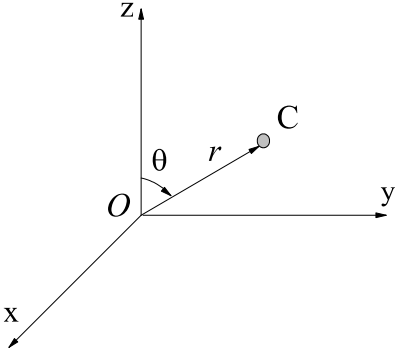}
\caption{Location of a cloud is determined by the radial distance $r$ and the polar angle $\theta$. We assume the orbital plane of the cloud is defined by $x$-axis and OC so that the inclination angle is $i$ and $  \widehat {XOC} = \psi $.}
\label{fig:f1}
\end{figure}

\section{General Formulation}
We consider a pressure-confined cloud with mass $m$ on its orbit in the gravitational field of a central object with mass $M$.  The gas cloud is treated as a  spherical solid body with a constant mass and no ablation. For simplicity, the intercloud gas is assumed to be static  which means the rotational and the radial velocities of the gas flow are neglected. A cloud is subject to the other forces such as a non-isotropic  radiation of the central accretion disc \citep{Liu11} and a drag force in the opposite direction of its orbital motion.  Another reasonable assumption is that the volume filling factor of the BLR cloud is low \cite[e.g.,][]{Rees89} which allows radiation pressure acts on each cloud directly and the shadowing effects are neglected. Dependence of the drag force on the relative velocity of the cloud and the surrounding gas is assumed to be linear. This functional form of the drag force is  known as Stokes' drag which gives frictional force on  a small spherical  objects in a laminar fluid. Such kind of gas flow occurs at low velocities and the nondimensional Reynolds number is a diagnostic to determine if the flow tends to be laminar or turbulent. 

 A primary uncertainty is that we do not yet know physical properties of the intercloud gas of a BLR region. If we use Stokes' drag relation, an estimate of the dynamics viscosity of the intercloud gas is needed. Although a BLR cloud is treated as a solid spherical object, it is actually a gaseous object which is possibly confined by an eternal agent like magnetic field. Thus, it is not clear if dynamics viscosity which is used for the motion a solid body subject to the Stokes drag is adequate for the motion of a gaseous pressure-confined cloud through an intercloud medium. We can not resolve this uncertainty unless further numerical simulations for the motion of a BLR cloud through an ambient gas are performed to obtain the drag force due to the interaction of the cloud with the intercloud. Thus, we approximate dynamic viscosity $\mu$ using a standard relation, i.e. $\mu =(1/2) \rho {\bar u} \lambda$, where $\rho $, ${\bar u}$ and $\lambda $ are the density of the gas,  the average molecular speed and  the mean free path, respectively. Also, we have $\lambda \approx 1/( d^2 \rho)$ where $d$ is the typical diameter of an atom or molecule. On the other hand, Reynolds number is defined as ${\rm Re}= \rho v L /\mu $, where $v$   is the mean velocity of the object relative to the gas and $L$  is a characteristic length. Thus, we have ${\rm Re} = 2 (v/{\bar u}) (L/\lambda )$. It is believed  that the intercloud gas with the number density $10^4 {\rm cm}^{-3}$ is at a temperature around or even larger than $10^8$ K \cite[e.g.,][]{whittle}. So, we have ${\bar u} \approx 2\times 10^6 {\rm m/s}$ and $\lambda \approx 10^{10} {\rm m}$. If we assume the mass of the central black hole is $10^6$ ${\rm M}_{\odot}$, then Keplerian velocity at 1 pc is approximately $6.5 \times 10^4 {\rm m/s}$. Obviously, the velocity of a cloud may reach several 1000 km/s in the vicinity of the central black hole. But we approximate the typical velocity of a cloud as $v \approx 6.5 \times 10^4 {\rm m/s} $. As for typical length scale, we approximate it as, say, ten times of the size of a BLR cloud, i.e. $L \approx 10^{13} cm$. Thus, we have ${\rm Re} \simeq 0.65$. Although our estimate of the Reynolds number provides  a justification for the assumption that the drag force has a linear dependence on velocity, we have high velocity clouds at the inner parts of the system which imply a much larger Reynolds number and then a quadratic relationship is more appropriate. But our analysis is still valid for the above mentioned conditions. Moreover, we did not consider the motion of the background gas. But if the BLR intercloud gas is rotating at a fraction of Keplerian velocity, then the relative velocity of a BLR cloud is reduced significantly which implies a smaller Reynolds number. These issues should be studied in future studies.  

Figure \ref{fig:f1} shows location of a cloud and our system of coordinates. Obviously, the angular momentum vector has a constant direction in space which implies the orbital plane has a fixed inclination angle $i$ with respect to the accretion disc, i.e. $x-y$ plane. Thus, location of the cloud can also be written in terms of angle $\angle XOC = \psi$ in the plane of motion and the radial distance $r$.

 Thus, equations of the orbital motion become
\begin{equation}
\ddot{r} - r\dot{\psi}^{2} = \frac{GM}{r^2} \left ( k |\sin \psi | - 1 \right ) - \frac{6\pi \mu R_{\rm cl}}{m} \dot{r} ,
\end{equation}
\begin{equation}
r \ddot{\psi} + 2 \dot{r} \dot{\psi} = - \frac{6\pi \mu R_{\rm cl}}{m} r \dot{\psi},
\end{equation}
where $\mu$ is the viscosity of intercloud gas and $R_{\rm cl}$ is the radius of the cloud. Moreover, the non-dimensional parameter $k$ is defined as
\begin{equation}
k = \frac{3l}{\mu_{\rm m} N_{\rm cl} \sigma_{\rm T}} \sin (i) .
\end{equation}
Here, $N_{\rm cl}$ is the column density of a spherical cloud and $\mu_{\rm m}$ is the mean molecular weight. The Eddington ratio is denoted by $l= L_{\rm a}/ L_{\rm edd}$ where $L_{\rm a}$ and $L_{\rm edd}$ are the luminosity of the central source and the Eddington luminosity, respectively. In the absence of drag force, the above equations reduce to the main equations of the previous studies \citep[e.g.,][]{plewa,khajenabi15}.

We note that in our model each cloud is in pressure equilibrium with the ambient gas. Then, radius of a cloud $R_{\rm cl}$ becomes in proportion to $P_{\rm gas}^{-1/3}$ where $P_{\rm gas}$ is the intercloud gas pressure.  So, behavior of the pressure of the intercloud gas with radius is an essential part of the present model for the pressure-confined clouds. It is generally assumed that the intercloud pressure distribution is proportional to a power-law function of the radial distance as $P_{\rm gas}  \propto r^{-s}$ where $s$ is an input parameter \citep[e.g.,][]{netzer2010}. However, this pressure profile may have a latitudinal dependence as it has been studied in detail by \cite{khajenabi15}. Here, we do not consider such a dependence for simplicity. Therefore, radius of each cloud varies according to its location within the intercloud gas as
\begin{equation}
R_{\rm cl} = R_{\rm cl0} \left ( \frac{r}{r_0}\right )^{s/3},
\end{equation}
where $r_0$ is a reference radial distance, say, the initial radial distance of the cloud, and $R_{\rm cl0}$ is the radius  of the cloud at $r_0$.   For our pressure-confined clouds, the column density becomes $N_{\rm cl} \propto P_{\rm gas}^{2/3}$ or $N_{\rm cl}=N_0 (r/r_0 )^{-2s/3}$ where $N_0$  is a constant column density. As for the   viscosity  of the gaseous intercloud medium, we prescribe it  as a power-law function of the radial distance, i.e.
\begin{equation}
\mu = \mu_0 \left( \frac{r}{r_0} \right)^\nu ,
\end{equation}
where $\nu$ is an input parameter and $\mu_0$ is the  viscosity of the intercloud gas at radius $r_0$. Thus, our non-dimensional parameter $k$ becomes $k=k_0 (r/ r_0 )^{2s/3}$ where $k_0 = 3l \sin (i) /{\mu_{\rm m}} N_0 \sigma_{\rm T}$. It is assumed that the temperature of a cloud does not change significantly during its orbital motion \citep{wang}.

Although our intention is not to study the confinement of BLR clouds by magnetic fields or other physical mechanisms, there are some observational and theoretical motivations for the power-law functions which are assumed for the intercloud gas pressure and dynamic viscosity. Some authors also used similar power-law functions to investigate orbital motion of BLR clouds and obtained the emission profiles that are consistent with observed emission spectra \citep[e.g.,][]{netzer2010}.  Although our knowledge of physical properties of the intercloud gas is still poor, an idea put forward by a few authors that the intercloud gas can be described using Advection-Dominated Accretion Flows (ADAFs), in which physical quantities are power-law functions of the radial distance \citep[e.g.,][]{krause12,khajenabi15}. If the temperature of the intercloud gas is at the virial temperature, then the exponent of the pressure distribution is $s=5/2$. But at the very inner parts of the system where the virial temperature is very high, the more appropriate value of the exponent is $s=3/2$ because of the efficiency of cooling. Because of uncertainties in the values of the exponents, we will do a parameter study for different values of the exponents. We also found that  for only one particular case with $s=0$ and $\nu=-2$, there is analytical solution for the orbit of the BLR clouds. Although we do not have enough observational information to confirm these particular values, the analytical solutions give a better insight of what we may expect for other values of the exponents.

In order to proceed, it is convenient to transform the main orbital equations into non-dimensional forms. In doing so, we use the reference radial distance $r_0$ to introduce a non-dimensional radial distance. Then, Keplerian velocity at this radial distance is written as $v_{\rm K}(r_0 ) = \sqrt{GM/ r_0 }$ and our time unit becomes $t_0 = r_0 / v_{\rm K}(r_0 )$. We now change  variables as $r \rightarrow r_0 r$ and $t \rightarrow t_0 t$. Thus,
\begin{equation}\label{eq:main1}
\ddot{r} - r \dot{\psi}^2 = \frac{1}{r^2} \left(k_0 r^{2s/3} |\sin\psi | - 1 \right) - \alpha r^{\nu + \frac{s}{3}} \dot{r},
\end{equation}
and
\begin{equation}\label{eq:main2}
r\ddot{\psi} + 2\dot{r}\dot{\psi} = -\alpha r^{\nu + \frac{s}{3} + 1} \dot{\psi},
\end{equation}
where $\alpha$ is a dimensionless drag coefficient, i.e.
\begin{equation}\label{eq:alpha}
\alpha = \frac{\Lambda}{\rm Re}.
\end{equation}
Here, Re is Reynolds number of the  initial condition and the non-dimensional parameter $\Lambda$ is
\begin{equation}\label{Lambda}
\Lambda = \frac{9}{2} (\frac{r_0}{R_{\rm cl0}}) (\frac{\rho_0}{\rho_{\rm cl0}})(\frac{u_0}{v_{\rm K}(r_0 )}).
\end{equation}
where $u_0$ is the typical velocity of  the cloud and $\rho_{\rm cl0}$ is the initial density of the cloud.  Moreover, the density of the atmosphere at the initial location of the cloud is denoted by $\rho_0$.

Equations (\ref{eq:main1}) and (\ref{eq:main2}) are our main equations for determining orbit of a cloud subject to a linear drag inside a gaseous medium with a power-law pressure distribution. It is unlikely to  obtain analytical solutions for all possible values of $s$ and $\nu$. We can obtain illustrative analytical solutions, however, for certain values of these input parameters as we will show below. 

\section{Analytical Solutions}

When variation of the intercloud gas pressure profile with the radial distance is not considered, we have $s=0$. In the absence of the drag, this particular case has already been studied analytically by \cite{plewa}. We now study this particular case, but considering the effects of the drag force. Thus, equation (\ref{eq:main2}) can be written as
\begin{equation}
\dot{L} + \alpha r^{\nu + 2} \dot{\psi} =0,
\end{equation} 
where $L=r^2 \dot{\psi}$ is the non-dimensional angular momentum per unit mass. If we set $\nu = -2$, this equation becomes integrable, i.e.
\begin{equation}
L=L_0 - \alpha\psi ,
\end{equation} 
where $L_0$ is a constant. Thus, magnitude of the angular momentum gradually decreases as a BLR cloud revolves around the central object. Total number of rotations before angular momentum becomes zero is $L_0 /(2\upi\alpha)$. 

Under substitution $u=1/r$ and changing the independent variable from time $t$ to angle $\psi$, we transform our main equations (\ref{eq:main1}) and (\ref{eq:main2}) into a differential equation which is integrable. Obviously, we have
\begin{equation}
\dot{r}=\frac{dr}{dt}=-\frac{1}{u^2}\frac{du}{dt}=-\frac{u'}{u^2}\dot{\psi},
\end{equation}
where the prime denotes differentiation with respect to the angle $\psi$.  Thus,
\begin{equation}
\ddot{r}=2\frac{u'^2}{u^3}\dot{\psi}^2-\frac{u''}{u^2}\dot{\psi}^2-\frac{u'}{u^2}\ddot{\psi}.
\end{equation}
Orbital equations (\ref{eq:main1}) and (\ref{eq:main2}) are re-written as
\begin{displaymath}
-\frac{u''}{u^2}\dot{\psi}^2 - \frac{u'}{u}(-2\frac{u'}{u^2}\dot{\psi}^2 + \frac{\ddot{\psi}}{u}+\alpha u \dot{\psi})-\frac{\dot{\psi}^2}{u^2}
\end{displaymath}
\begin{equation}\label{eq:mu1}
=u^2(k_0 |\sin\psi|-1),
\end{equation}
and
\begin{equation}\label{eq:mu2}
-2\frac{u'}{u^2}\dot{\psi}^2 + \frac{\ddot{\psi}}{u}+\alpha u \dot{\psi}=0.
\end{equation}
According to equation (\ref{eq:mu2}), the coefficient of the ratio $u'/u$ in equation (\ref{eq:mu1}) vanishes. Thus,
\begin{equation}
u'' + u = \frac{u^4}{\dot{\psi}^2} (1-k_0 |\sin\psi|).
\end{equation}
We also showed that $\dot{\psi}=u^2 (L_0 - \alpha\psi )$, and so the above equation becomes

\begin{equation}\label{eq:pfinal}
\frac{d^2 u}{d\psi^2} + u = \frac{1-k_0 |\sin\psi |}{(L_0 - \alpha\psi )^2}.
\end{equation} 
If we introduce a new variable $z=(L_0 - \alpha\psi )/\alpha$, this equation can be simplified further and it becomes
\begin{equation}\label{eq:final}
\frac{d^2 u}{dz^2} + u = \frac{1-k_0 |\sin (\frac{L_0}{\alpha} -z)|}{\alpha^2 z^2}.
\end{equation}

If we neglect the drag force, then equation (\ref{eq:pfinal}) reduces to equation (7) in \cite{plewa} who studied orbits of BLR clouds without a resistive force. They obtained an analytical solution and a condition for having bound orbits. Their analytical solution is written as
\begin{equation}
u(\psi ) = 1+\frac{k_0}{2} \psi\cos\psi + A \sin\psi + B \cos\psi ,
\end{equation}
for $0\leq \psi \leq \pi$ and similarly for $\pi < \psi \leq 2\pi$ , we have
\begin{equation}
u(\psi ) = 1- \frac{k_0}{2} \psi\cos\psi + (A+k_0 ) \sin\psi + (B+k_0 \pi ) \cos\psi ,
\end{equation}
where $A$ and $B$ are arbitrary constants and both  $u$ and $du/d\psi$ are continuous  at $\psi = \pi$. Having the above analytical solutions, \cite{plewa} investigated orbits of the BLR clouds. Obviously, in the presence of drag force, the orbits would not be bound. Interestingly, we can find an analytical solution when the drag force is considered, too.  

Moreover, if we neglect non-isotropic force due to the central radiation field,  equation (\ref{eq:final}) reduces to equation (12) in \cite{mitt}  which is integrable  and the solution is written in terms of the sine and cosine integrals. We can also present solutions of equation (\ref{eq:final}) in terms of these integrals. For $2n\pi < \psi < (2n+1) \pi$ ($n=0,1,2,...$), the general solution of equation (\ref{eq:final}) becomes
\begin{displaymath}
u(z)=
\end{displaymath}
\begin{displaymath}
A_{2n} \cos (z) + B_{2n} \sin (z) - \frac{1}{\alpha^2}\left [ \sin (z) {\rm Si}(z) + \cos (z) {\rm Ci}(z)\right ]
\end{displaymath}
\begin{equation}
-\frac{k_0}{\alpha^2} \left [ \cos (\frac{L_0}{\alpha} +z) {\rm Si}(2z) \sin (\frac{L_0}{\alpha} +z) {\rm Ci}(2z) \right ]
\end{equation}
and for $(2n+1)\pi < \psi < (2n+2) \pi$, the solution is
\begin{displaymath}
u(z)=
\end{displaymath}
\begin{displaymath}
C_{2n} \cos (z) + D_{2n} \sin (z) + \frac{1}{\alpha^2}\left [ \sin (z) {\rm Si}(z) + \cos (z) {\rm Ci}(z)\right ]
\end{displaymath}
\begin{equation}
+\frac{k_0}{\alpha^2} \left [ \cos (\frac{L_0}{\alpha} +z) {\rm Si}(2z) \sin (\frac{L_0}{\alpha} +z) {\rm Ci}(2z) \right ]
\end{equation}
where $A_{2n}$, $B_{2n}$, $C_{2n}$ and $D_{2n}$ are arbitrary constants and the continuity of $u$ and $du/dz$ at $\psi = 2n\pi$ and $\psi = (2n+1)\pi$ for a given $n$ imply that not all these parameters are independent.

Our analytical solutions clearly demonstrate decay of orbits because of the presence of the drag force. But shape of the orbits and the  time for a BLR cloud to travel from an initial radius to the center depend on the initial conditions and the input parameters. 
 
\section{Analysis}
\begin{figure}%[tb]
\includegraphics[scale=0.65]{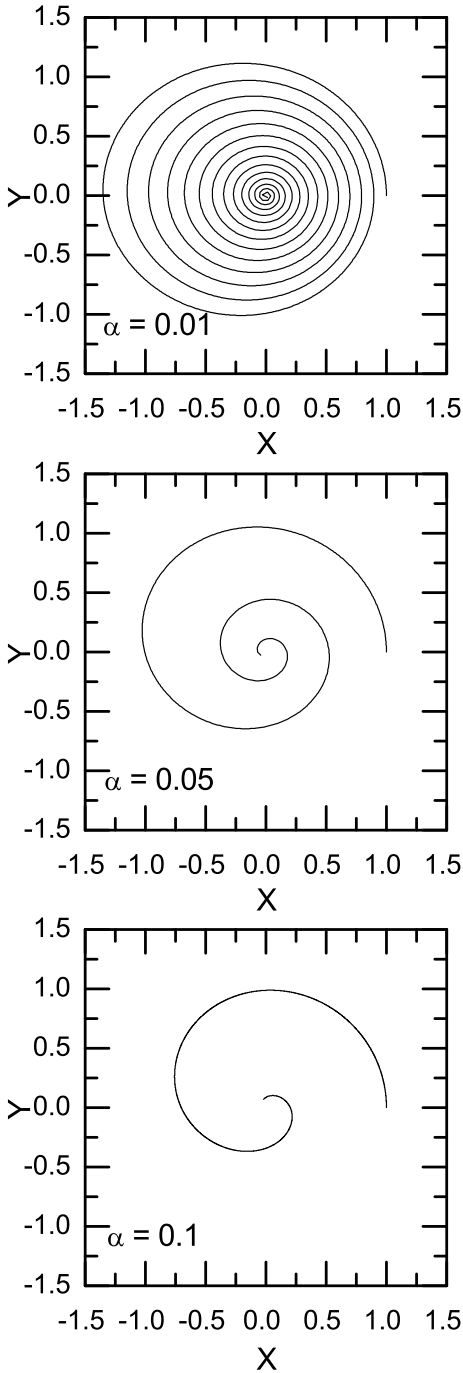}
\caption{Orbit of a BLR cloud when its journey starts from the initial location $r=1$ and the input parameters are $s=0$, $\nu =-2$, $k_0 =0.1$ and $L_0 = 1$. Initial conditions are $\dot{r}(t=0)=0$, $\psi (t=0) =0$ and $\dot{\psi}(t=0)=1$. Each plot is labeled by the corresponding value of $\alpha$. } %% no full stop at the end of caption
\label{fig:2}
\end{figure}
%% Math 
%
%
First, we study orbital shape of a BLR cloud in the plane of motion for a particular case with $s=0$, $\nu=-2$, $L_0 =1$ and  $k_0 =0.1$ in Figure \ref{fig:2}. We numerically solve the orbital equations and our analytical solutions for this particular case are used to check the accuracy of our numerical solutions. Differences between our analytical and numerical solutions are found to be very negligible. In this figure and all subsequent figures, we consider the same initial conditions, i.e. $r(t=0)=1$,  $\dot{r}(t=0)=0$, $\psi (t=0) =0$ and $\dot{\psi}(t=0)=1$. Moreover, the inclination angle of plane of motion $i$ is fixed so that X axis coincides x axis and Y axis make angle $i$ with y axis in $y-z$ plane (Figure \ref{fig:f1}). We also explored other initial conditions, but the results are qualitatively similar to what  we are reporting here for a few illustrative cases.  Figure \ref{fig:2} shows shape of the orbits for three values of the dimensionless resistive  coefficient $\alpha$. It is apparent that the decay of the orbit is much faster as $\alpha$ goes from $0.01$ to $0.1$. We can also calculate the  time for a BLR cloud to travel  from its initial location to the center which is defined as the time-of-flight $t_{f}$. Considering orbits in Figure \ref{fig:2} we can say that  time-of-flight is significantly reduced with increasing the coefficient $\alpha$.  

\begin{figure}%[tb]
\includegraphics[scale=0.5]{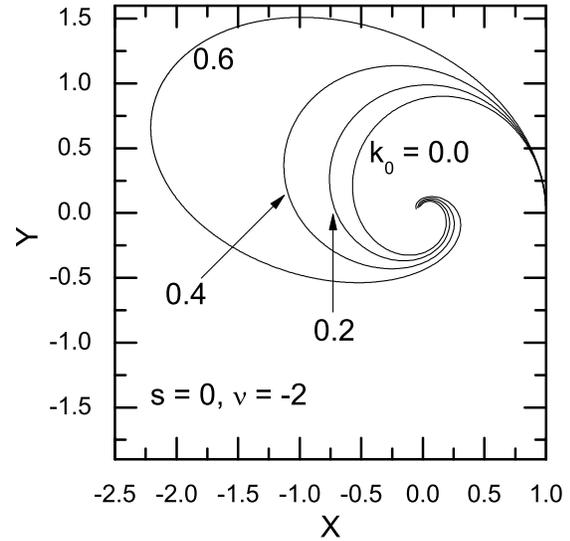}
\caption{Orbits of a BLR clouds when all the input parameters and the initial conditions are similar to Figure \ref{fig:2}, except that $k_0$ which varies from 0 to 0.6 in order to illustrate role of non-isotropic radiation in shape of the orbits. Each orbit is labeled by the corresponding value of $k_0$.} %% no full stop at the end of caption
\label{fig:3}
\end{figure}

In Figure \ref{fig:3}, the effect of the non-isotropic radiation force is studied for a configuration where $s=0$, $\nu =-2$ and $\alpha=0.1$. A case with $k_0 =0$ corresponds to a system without a central source of the radiation. We see that the orbits become more elongated as the parameter $k_0$ increases because of the non-isotropic nature of the central radiation field. Thus, time-of-flight $t_{f}$ increases as $k_0 $ increases  from 0 to 0.6.

\begin{figure}%[tb]
\includegraphics[scale=0.5]{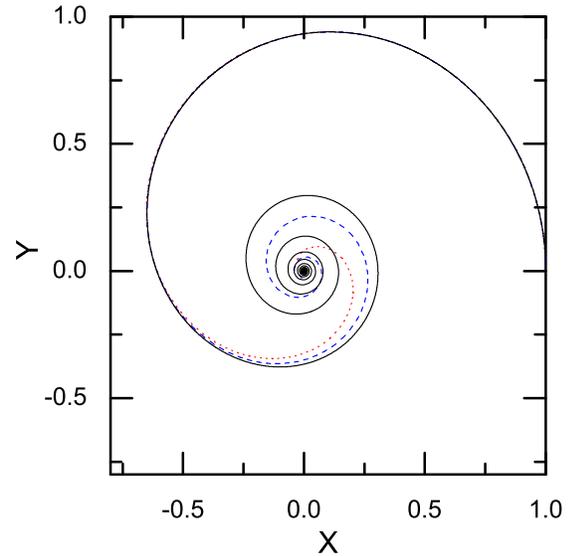}
\caption{Orbits of a BLR cloud for different values of $s$, i.e. $s=2.5$ (solid-black), $s=1.5$ (dashed-blue), $s=0$ (dotted-red).  The other input parameters are $\nu =-2$, $\alpha =0.1$, $k_0 = 0.1$, $\dot{r}(t=0)=0$, $\psi (t=0) =0$ and $\dot{\psi}(t=0)=1$.} %% no full stop at the end of caption
\label{fig:4}
\end{figure}
We prescribed the background pressure as a simple power-law function of the radius with the exponent $s$. For our pressure-confined clouds, the exact mechanisms of the confinement are not considered as long as the pressure profile is prescribed as a power-law function of the radius. Figure \ref{fig:4} shows orbits of a BLR cloud for different values of  the exponent $s$ while keeping all the rest of  the input parameters unchanged. Solid, dashed and dotted lines are corresponding to $s=2.5, 1.5$ and 0, respectively. This Figure indicates  that decay of the orbits depends on the exponent $s$ so that  time-of-flight $t_f$ becomes longer as the value of $s$ increases from 0 to 2.5.  Drag force is directly proportional to the radius of a cloud which is a function of its location because of pressure-confinement condition. Thus,  drag force also depends on the radial distance so that a smaller value for $s$ implies a larger drag force. For this reason, when we have $s=0$, a BLR cloud very quickly falls onto the central object in comparison to the situations with the larger values for $s$.

\begin{figure}%[tb]
\includegraphics[scale=0.5]{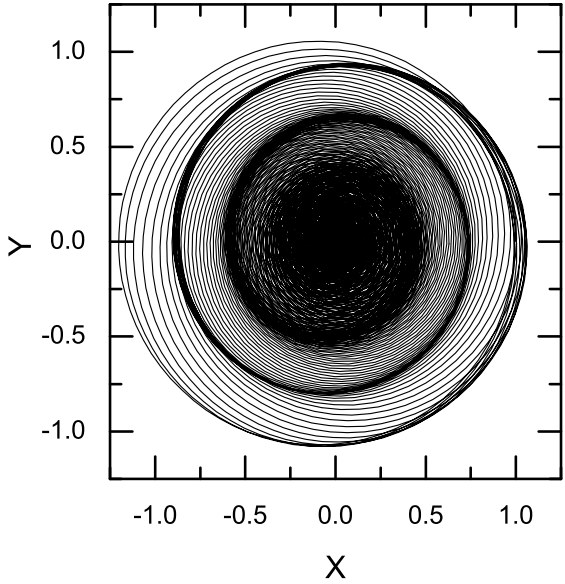}
\caption{Orbit of a BLR cloud when the dimensionless drag coefficient is very small, i.e. $\alpha =0.001$. Input parameters are $\nu =-2$,  $k_0 = 0.1$ and $s=2.5$ and the initial conditions are same as Figure \ref{fig:2}. } %% no full stop at the end of caption
\label{fig:5}
\end{figure}

When the drag coefficient is small, a cloud will decay very slowly but the non-isotropic radiation force causes the orientation of the orbit changes smoothly so that a cloud on its orbit may come back to same region for several times. In other words, presence's probability of a cloud at specific regions is larger than other parts depending on the input parameters and under certain circumstances. As an illustrative example of this case, orbit of a BLR cloud for a small drag coefficient but in the presence of non-isotropic radiation field is shown in Figure \ref{fig:5}. We see that a cloud on its orbit repeats certain locations so that spiral patterns will emerge as a whole. If we start with an ensemble of identical clouds and the same initial conditions, the spatial distribution of the clouds as they  orbit will resemble to these spiral patterns. Although emergence of such a configuration is extremely unlikely and rather of academic interest, existence of these complicated orbits show richness and  diversity of orbits when the drag force and the non-isotropic radiation field are taking into account in addition to the central gravitational force.

\begin{figure}%[tb]
\includegraphics[scale=0.65]{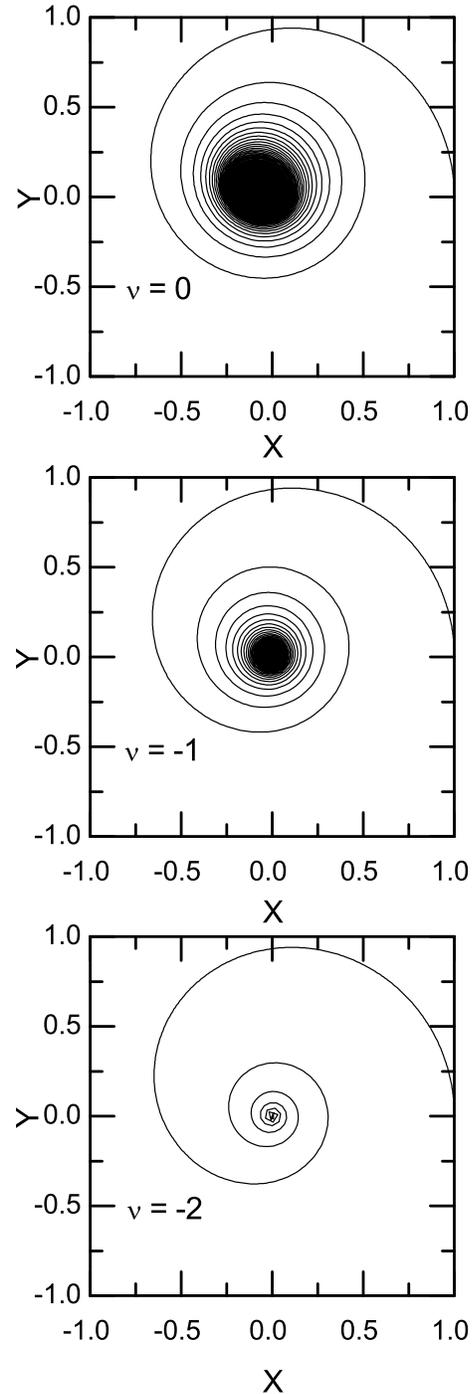}
\caption{Orbit of a BLR cloud when we have $\alpha =0.1$, $k_0 = 0.1$ and $s = 2.5$. Each plot is labeled by the corresponding value of $\nu$. Again initial conditions are same as Figure \ref{fig:2}.} %% no full stop at the end of caption
\label{fig:6}
\end{figure}

\begin{figure}%[tb]
\includegraphics[scale=0.5]{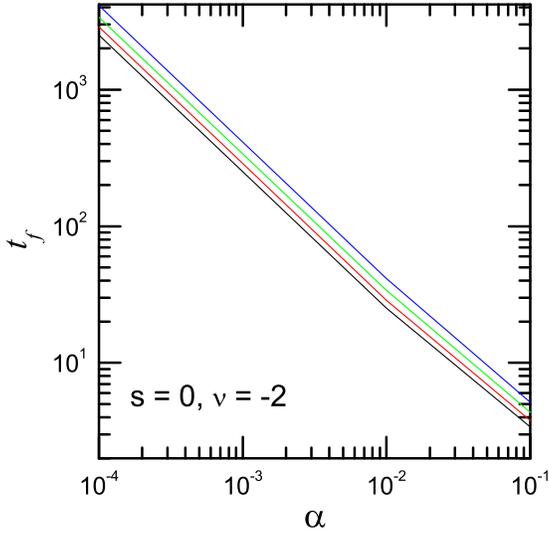}
\caption{Time-of-flight $t_{f}$ as a function of the dimensionless drag coefficient $\alpha$ when the input parameters are $s=0$ and $\nu=-2$ and the initial conditions are same as Figure \ref{fig:2}. Curves from below to top are corresponding to $k_0$ equal to 0, 0.1, 0.2 and 0.3, respectively.} %% no full stop at the end of caption
\label{fig:7}
\end{figure}

In our model, the viscosity  is also prescribed as a power-law function of the radius with the exponent $\nu$. We examine effect of varying  parameter $\nu$ on the shape of  orbits in Figure \ref{fig:6}. Here, we assume that $\alpha =0.1$, $k_0 = 0.1$, and $s=2.5$, but different values of $\nu$ are considered. As the viscosity becomes steeper with the radius, then drag force becomes stronger which result in shorter time-of-flight. For other input parameters,  dependence of the orbits on the variations of $\nu$ is found qualitatively similar to Figure \ref{fig:6}. 

Profile of $t_f$ as a function $\alpha $ is shown for a particular case in Figure \ref{fig:7}. The initial conditions are similar to the previous Figures and we have $s=0$ and $\nu=-2$. Different values for $k_0$ are assumed. Curves from below to top are corresponding to $k_0 = 0$, 0.1, 0.2, and 0.3, respectively. This Figure shows that $t_f \propto \alpha^{-1}$ and the time-of-flight becomes longer with increasing $k_0$. 

\section{Discussion}

Our knowledge about origin  of BLR clouds is not satisfactory. We do not know if these clouds are constantly created  or they have formed more or less simultaneously some  time in the past.  In fact, there is an ongoing debate about if these clouds are stable structures or else they are transient objects which are constantly created and destroyed in a turbulent medium. It is well-known that magnetic fields play an important role in confining the clouds and if their effects are neglected the clouds would not be stable and very soon after their formation they will expand and disperse \citep[e.g.,][]{Rees}. Numerical simulations of  clumps in a radiative medium under restrictive conditions such as neglecting magnetic effects show that clouds are not able to move around because of the force due to the radiation pressure. On the other hand, just recently \cite{McCourt} showed that if magnetic fields and their role in confinement of the clouds are considered,  the clouds not only become stable objects but also they can move in their orbits. As for the G2 cloud in the Galactic center, however, they predict that drag forces can lead to deviations from Keplerian profile. Thus, this issue deserves further investigations in particular regarding magnetic confinement and possible role of magnetic fields in modifying structure of the intercloud gas.  But if the BLR clouds do really exist, they should be long-lived objects because their dominant role in the emission from such systems. There is also a possibility that formation and destruction of the clouds are so fast that only on the basis of the current received emission we can not distinguish  weather they are long-lived stable objects or not, unless  more theoretical and observational diagnostics are proposed.

Irrespective of the stability issue and the exact confinement mechanism,  we assume that the clouds are stable long-lived objects which have formed some time in the past. Because of the drag force, all clouds will eventually fall onto the central black hole or they may destroy at the very inner parts due to gravitational tide or intense central radiation. In any case,   time-of-flight $t_{\rm f}$ of a BLR cloud which starts its life at the outer parts of system should be longer than the life-time of the whole system, i.e. $\tau $. If the time-scale $t_{\rm f}$ becomes shorter than the life-time of the whole system, one can expect all BLR cloud will eventually fall onto the central region in the absence of any continuous mechanism for creation of BLR clouds. In other words, the system will be depleted of BLR clouds in this case. But emission of system is interpreted in terms of existence of these cold clouds. So, there is a contradictory situation regarding to the existence of BLR clouds if time-of-flight becomes shorter than the life-time of the system unless we propose robust mechanisms for continually  creating BLR clouds so that even they fall onto the central object, the creation mechanisms lead to the formation of new clouds. The net effect is a clumpy system, though some of the clouds may be destroyed  because of falling onto central object. In our orbital analysis, time-of-flight $t_{\rm f}$ is very sensitive to the input parameter, in particular dimensionless drag coefficient $\alpha$ has a vital role so that we can have very long or very short $t_{\rm f}$ depending on this coefficient. 

If we set the life-time of the system, then $t_{\rm f}$ should be longer than this life-time in order to have a clumpy system in the absence of creation mechanism. This constraint will gives us a constraint on the value of $\alpha$ which depends on local properties of intercloud gas. On the other hand, if we can constrain the input parameters including $\alpha$ independently, then we can compare $t_{\rm f}$ to the life-time of the system. As we discussed above, a case with $t_{\rm f}$ shorter than the life-time of the system, implies existence of mechanisms for continuous creation of BLR clouds. But if $t_{\rm f}$ becomes longer than the life-time of the system such a formation mechanism may or may not exist. 

For a wide range of the input parameters, we found that $t_{\rm f} \simeq t_{0}(\nu,s)/\alpha$ where the parameter $t_0$ depends on the exponents $\nu$ and $s$ (Figure \ref{fig:7}). Now, if we have $t_{\rm f} < \tau $, then there might be a continuous mechanism for replenishment of BLR clouds. Thus, existence of such a mechanism implies a lower limit for the age of the whole system, i.e. $\tau_{\rm min} \simeq t_{0}(\nu,s)/\alpha $. If the age of system is older than $\tau_{\rm min}$, we can conclude that BLR clouds are continuously  created and possibly destroyed. Using equations (\ref{eq:alpha}) and (\ref{Lambda}), we then have $\tau_{\rm min} = (2 {\rm Re} /9)  t_{0} (\nu , s) (R_{\rm cl0}/r_0) ( \rho_{\rm cl0}/\rho_0 )$ where we have assumed that $u_0 \simeq v_{\rm K}(r_0 )$. Observations suggest a BLR region has a typical size of $\sim 0.01-1 {\rm pc}$ which contains a large number of clouds with a very low filling factor ($\sim 10^{-8}$) and typical sizes $\sim 10^{12}-10^{14} {\rm cm}$ and mass $\sim 10^{-8} {\rm M_{\rm\odot}}$ \citep[][]{Rees89,marconi,plewa}. Now, we can estimate $\tau_{\rm min}$ if we set ${\rm Re} \simeq 1$ and assume that a cloud starts its journey from the outer parts of the BLR region, i.e. $r_0 = 0.01 {\rm pc}$. Also, we have $R_{\rm cl0} = 10^{14} {\rm cm}$. Thus, we obtain $\tau_{\rm min} = 7.2 \times 10^{-4} ( \rho_{\rm cl0}/\rho_0 ) t_{0} (\nu , s) $. A larger density contrast $\rho_{\rm cl0}/\rho_0$ implies a smaller cloud if the mass is kept fixed. If we set $\rho_{\rm cl0}\simeq 10^{10} {\rm cm}^{-3}$ and $\rho_{0} \simeq 10^{4} {\rm cm}^{-3}$ \citep[e.g.,][]{whittle}, then we have $\tau_{\rm min} =720 t_{0}(\nu, s)$. Although value of $t_0 (\nu , s)$ depends on the exponents $\nu$ and $s$, its variation with these parameters is not very significant so that we generally found that $t_{0}(\nu, s) <1$. Thus, if we assume $ t_{0}(\nu, s) \simeq 1$, then  $\tau_{\rm min} =720$ or $\tau_{\rm min} \simeq 15\times 10^6 {\rm yr}$ for a cloudy BLR region around a central black hole with mass $10^8 {\rm M}_{\odot}$. But it is unlikely that a BLR region to be younger than this time scale. In estimating $\tau_{\rm min}$, we considered the input parameters so that the largest possible value for this time-scale is obtained. In other words, the actual value of $\tau_{\rm min}$ is probably much shorter than 15 million years unless the clouds become very compact with a very large density contrast which is not supported by observations.  

Line profiles of the BLR clouds based on their orbital motion has been calculated by some authors in the absence of the drag force \citep[e.g.,][]{netzer2010}. Considering the importance of  the drag force in changing the orbits, as we showed in this study, future work is also needed  to calculate line profiles of the BLR clouds according to our orbital solutions. Moreover, we assumed that the drag force is  a linear function of the velocity. Although it enabled us to obtain analytical solution for a certain set of the input parameters,  the next step is to obtain orbits of BLR clouds when drag force is proportional to the velocity square of a cloud.

\section{Conclusion}
We investigated orbital motion of the BLR clouds which are subject to the gravitational force of the central black hole, a non-isotropic force due to an accretion disc and a resistive force in proportion to the relative velocity of the cloud and the intercloud gas. Although direction of the angular momentum vector does not change, its magnitude reduces because of the effect of the drag force and therefore this force inevitably causes the clouds to spiral downward. Assuming that the physical properties of the intercloud medium like its pressure and the viscosity  are described as power-law functions of the radial distance, we found that the orbit of a cloud and its time-of-flight  sensitively depend on the non-dimensional parameter $\alpha$ which is a function of the properties of gaseous component at the initial location of the  cloud. Our main conclusion is that most of the BLR clouds will fall onto the central black hole because their time-of-flight is shorter than life-time of whole system. This result is more or less independent of the exponents of the pressure distribution and dynamic viscosity. A constant replenishment mechanism for the BLR clouds is needed when the drag force is considered.

\section*{Acknowledgments}
This work has been supported financially by Research Institute for Astronomy \& Astrophysics of Maragha (RIAAM) under research project No. 1/3720-60. I am very grateful to referee for his/her report which greatly improved quality of the paper. 

\bibliographystyle{mn2e}
\bibliography{reference}

\bsp

\label{lastpage}

\end{document}